\documentclass[prl,twocolumn,showpacs,amsfonts,amsmath,floatfix]{revtex4}
\usepackage{graphicx}
\usepackage{psfrag}
\newcommand{\Journal}[4]{#1 {\bf #2}, #3 (#4)}
\newcommand{\PR}{Phys. Rev.}
\newcommand{\PRL}{Phys. Rev. Lett.}
\newcommand{\PRA}{Phys. Rev. A}
\newcommand{\PRB}{Phys. Rev. B}
\newcommand{\JMP}{J. Math. Phys.}

\newcommand{\Science}{Science}

\newcommand{\PLA}{Phys. Lett. A}
\begin{document}
\title {Anyon-fermion mapping and applications to ultracold gases in tight 
waveguides}
\author{M. D. Girardeau}
\email{girardeau@optics.arizona.edu}
\affiliation{College of Optical
Sciences, University of Arizona,
Tucson, AZ 85721, USA}
%
\begin{abstract}
The Fermi-Bose mapping method for one-dimensional (1D) Bose 
and Fermi gases with zero-range interactions is generalized to an anyon-fermion
mapping and applied to exact solution of several models of ultracold gases 
with anyonic exchange symmetry in tight waveguides: anyonic Calogero-Sutherland
model, anyons with point hard core interaction (``anyonic TG gas''), and 
spin-aligned anyon gas with infinite zero-range odd-wave attractions 
(``anyonic FTG gas'').  It is proved that for even $N\ge 4$ there are 
states of the anyonic FTG gas on a ring, with anyonic phase slips which
are odd integral multiples of $\pi/(N-1)$, of energy lower than that of the 
corresponding fermionic ground state. A generalization to a spinor Fermi gas
state with anyonic symmetry under purely spatial exchange enables energy 
lowering by the same mechanism.
\end{abstract}
\maketitle
The Fermi-Bose (FB) mapping method was introduced in 1960 \cite{Gir60} and
used to obtain the exact $N$-particle ground and excited states of a 1D
gas of impenetrable point bosons [now known as the Tonks-Girardeau (TG)
gas], which has recently become the subject of extensive theoretical 
and experimental investigations 
because of the novelty and experimental realizability \cite{Par04,Kin04}
of ultracold gases in tight atom waveguides with strong correlations 
induced by Feshbach resonance tuning \cite{Rob01} of the effective 1D 
interactions to very large values via confined-induced resonances 
\cite{Ols98,BerMooOls03,GraBlu04}. It is now known 
\cite{CheShi98,GirOls03,GirNguOls04,GraBlu04} that the FB mapping
is of much greater generality; when supplemented by an inversion and sign 
change of the coupling constant, it provides a mapping between
the $N$-body energy eigenstates of a 1D Bose gas with delta-function 
interactions of any strength [Lieb-Liniger (LL) gas \cite{LieLin63}] and those 
of a spin-aligned Fermi gas. For a recent review see \cite{YukGir05}. 

In the three-dimensional world experimental evidence supports the
\emph{symmetrization postulate} (SP), according to which wave functions of
identical particles are either completely symmetric (Bose) or antisymmetric 
(Fermi) under permutations of the particle coordinates \cite{MesGre64}.
The spin-statistics theorem entertains no possibilities other than bosons or 
fermions, excluding more complicated permutation symmetries from the start
\emph{by hypothesis} \cite{StrWig64}. There is a trivial 
 ``proof'' \cite{Cor51} of SP in some textbooks, but it is incorrect;
by generalizing the original FB mapping \cite{Gir60} I identified 
the logical error in this ``proof'' and pointed out that for a 1D system of 
identical particles with hard-core interactions, wave functions which are
neither completely symmetric (Bose) nor completely antisymmetric (Fermi) 
are physically allowed \cite{Gir65}. Leinaas and Myrheim \cite{LeiMyr77} 
generalized and rigorized this approach, proved SP in 3D, and showed that 
permutation symmetries interpolating continuously between bosons and fermions 
are not excluded in 1D and 2D. In recent years low-dimensional systems with 
anyonic symmetry have  
found application both in relativistic particle physics and condensed matter 
(quantized fractional Hall effect \cite{Lau83,Hal84,Cam05}, 
anyonic superconductivity \cite{Lau88,Wil90}, rotating ultracold gases 
\cite{WilGun00,CooWil99,Par01}, Aharonov-Bohm and Aharonov-Casher effects
and persistent current in 1D mesoscopic rings \cite{ZhuWan96}, 
and quantum-knot computation
\cite{Sar05}).

Most of these applications involve systems which are essentially 
two-dimensional, but anyonic symmetry can also occur in 1D 
\cite{ZhuWan96,Kun99}, where anyonic exchange symmetry
leads to strong short-range correlations. An anyonic generalization of the FB 
mapping will be used herein to obtain exact solutions for several models of 
ultracold gases with 1D anyonic exchange symmetry in tight waveguides: anyonic 
Calogero-Sutherland (CS) model, anyonic TG gas, and anyonic FTG gas.  The TG 
gas has already been physically realized via Feshbach
resonance tuning of the effective 1D interactions in Bose gases in tight 
waveguides \cite{Par04,Kin04}, the fermionic TG (FTG) gas may be realizable by 
the same mechanism, and it has recently been pointed out
\cite{Yu06} that the CS model of bosons with inverse square repulsive
potential should also be realizable. 

{\it 1D Anyon fields, anyonic symmetry, and zero-range interactions:} 
In 1D particles can only physically exchange positions by passing through each
other, so exchange symmetry is inseparable from short-range 
interactions. This is the origin of
the FB mapping from impenetrable point bosons (TG gas) to noninteracting 
spin-aligned fermions \cite{Gir60}, and more generally of the Fermi-Bose 
duality in 1D \cite{KorBogIze93}. Several similar but nonequivalent
definitions of 1D anyons appear in the literature. Kundu \cite{Kun99}
defines anyon field operators $\hat{\psi}_A(x)$ in terms of Bose operators  
$\hat{\psi}_B(x)$ by 
$\hat{\psi}_A(x)=e^{-i\theta\int_{-\infty}^{x}dx'\hat{\rho}(x')}
\hat{\psi}_B(x)$  
where $\hat{\rho}(x)=\hat{\psi}_B^{\dagger}(x)\hat{\psi}_B(x)
=\hat{\psi}_A^{\dagger}(x)\hat{\psi}_A(x)$ is the number density operator.
These satisfy exchange relations 
$\hat{\psi}_A(x)\hat{\psi}_A^{\dagger}(x')=\delta(x-x')
+e^{-i\theta\epsilon(x-x')}\hat{\psi}_A^{\dagger}(x')
\hat{\psi}_A(x)$ and
$\hat{\psi}_A(x)\hat{\psi}_A(x')=
e^{i\theta\epsilon(x-x')}\hat{\psi}_A(x')\hat{\psi}_A(x)$
where $\epsilon(x)=+1\ (-1)$ for $x>0\ (x<0)$, and $\epsilon(0)=0$.  
In this paper $x$ and $x'$ are 1D coordinates, bearing in mind
the effectively 1D dynamics of ultracold gases in wave guides with transverse
trapping so tight that the transverse excitation energy quantum exceeds
the available longitudinal zero-point energy \cite{Ols98}. Kundu carried out a 
formal Bethe ansatz solution for the $N$-body energy eigenstates of such a 
system of anyons starting from a contact condition of LL form \cite{LieLin63}. 
However, an attempt to generate these contact conditions as a zero-range limit 
of boundary conditions at the edges of a finite-range
interaction potential leads to contradictions, and application of the kinetic 
energy operator to a wave function with contact discontinuities generates 
singularities which can only be cancelled by highly singular and ill-defined 
interactions \cite{Kun99}. 

A different definition closely related to fermions eliminates these 
difficulties. Define the anyon field annihilation operator $\hat{\psi}_A(x)$ in
terms of the Fermi field operator $\hat{\psi}_F(x)$ by
$\hat{\psi}_A(x)=e^{-i\theta\int_{-\infty}^{x}dx'\hat{\rho}(x')}
\hat{\psi}_F(x)$  
where $\hat{\rho}(x)=\hat{\psi}_F^{\dagger}(x)\hat{\psi}_F(x)
=\hat{\psi}_A^{\dagger}(x)\hat{\psi}_A(x)$ is the number density operator.
Then 
\begin{eqnarray}\label{exchange}
& &\hat{\psi}_A(x)\hat{\psi}_A^{\dagger}(x')
+e^{-i\theta\epsilon(x-x')}\hat{\psi}_A^{\dagger}(x')
\hat{\psi}_A(x)
=\delta(x-x')\nonumber\\
& &\hat{\psi}_A(x)\hat{\psi}_A(x')
+e^{i\theta\epsilon(x-x')}\hat{\psi}_A(x')\hat{\psi}_A(x)=0\ .
\end{eqnarray}
Then the exclusion principle $\hat{\psi}^2(x)=[\hat{\psi}^{\dagger}(x)]^2=0$\ 
follows from $\epsilon(x)=0$, ensuring that anyonic 
phase discontinuities only occur at collisional nodes of
the wave functions. At such zeros the phase is undefined, allowing
phase slips consistent with those required by anyonic exchange
symmetry. 

$N$-particle anyon wave functions $\Psi_A$ are the
amplitudes in an $N$-anyon
Fock state 
$|\Psi_A\rangle=(N!)^{-\frac{1}{2}}\int dx_1\cdots dx_N
\Psi_A(x_1,\cdots,x_N)\hat{\psi}_A^{\dagger}(x_1)\cdots
\hat{\psi}_A^{\dagger}(x_N)|0\rangle$. 
Using 
Eq. (\ref{exchange}) to exchange $\hat{\psi}_A^{\dagger}(x_j)$ and
$\hat{\psi}_A^{\dagger}(x_{j+1})$ and interchanging the names of these
integration variables, one proves that
$\Psi_A(\cdots,x_j,x_{j+1},\cdots)=-e^{-i\theta\epsilon(x_{j+1}-x_j)}
\Psi_A(\cdots,x_{j+1},x_j,\cdots)$, i.e., there is a sign change plus a phase 
slip $e^{i\theta}$ ($e^{-i\theta}$) when any particle passes its neighbor to 
the right (left). Iterating this one proves
\begin{widetext}
\begin{equation}\label{symmetry}
\Psi_A(x_1,\cdots,x_j,\cdots,x_k,\cdots,x_N)
=-e^{i\theta[\sum_{\ell
=j+1}^k\epsilon(x_{\ell}-x_j)+\sum_{\ell=j+1}^{k-1}
\epsilon(x_k-x_{\ell})]}\Psi_A(x_1,\cdots,x_k,\cdots,x_j,\cdots,x_N)
\end{equation}
\end{widetext}
which is similar to Kundu's Eq. (11) \cite{Kun99}, but has the very
important difference that these wave functions satisfy the exclusion
principle, i.e., $\Psi_A$ vanishes when $x_j=x_k$ for all $j\ne k$.
Anyonic symmetry of this type may have applications to electrons in
mesoscopic rings \cite{ZhuWan96} as well as to ultracold gases in the 1D 
regime. $\Psi_A$ cannot satisfy periodic boundary 
conditions with periodicity length $L$ unless $\theta$ is a multiple of 
$2\pi/(N-1)$; instead, it satisfies \emph{twisted boundary conditions}
$\Psi_A(x_1,\cdots,x_j\pm L,\cdots,x_N)
=(-1)^{N-1}e^{\mp i(N-1)\theta}\Psi_A(x_1,\cdots,x_j,\cdots,x_N)$. 
If one places the particles on a ring of circumference $L$
and requires that wave functions be single-valued, then the only values
of $\theta$ allowed if $N$ is odd are integral multiples of $2\pi/(N-1)$,
and if $N$ is even the only allowed values are odd integral multiples of 
$\pi/(N-1)$. As $N\to\infty$ this set of allowed $\theta$ values becomes 
dense.

{\it Anyon-fermion mapping:} Define an anyon mapping
function $A_{\theta}$ by
$A_\theta(x_{1},\cdots,x_{N})=\prod_{1\le j<k\le N}
e^{\frac{1}{2}i\theta\epsilon(x_{jk})}$ where $x_{jk}=x_j-x_k$.
This generalizes the original FB mapping \cite{Gir60,Gir65}, to which it
reduces, apart from an irrelevant constant factor, when $\theta=\pi$.  
Define $\Psi_F$ by 
$\Psi_F(x_1,\cdots,x_N)
=A_{\theta}\Psi_A(x_1,\cdots,x_N)$
where $\Psi_A$ is an $N$-anyon wave function satisfying Eq. (\ref{symmetry}).
Then $\Psi_F$ is 
totally antisymmetric (fermionic). Conversely, if one defines $\Psi_A$ by 
$\Psi_A(x_1,\cdots,x_N)
=A_{-\theta}\Psi_F(x_1,\cdots,x_N)$
where $\Psi_F$ is fermionic, then $\Psi_A$ satisfies (\ref{symmetry}).
Finally, if one defines $\Psi_B$ by $\Psi_B=A_{\theta+\pi}\Psi_A$ where
$\Psi_A$ satisfies (\ref{symmetry}), then $\Psi_B$ will be completely
symmetric (bosonic). 

{\it Anyonic CS gas:} Consider bosons or fermions on a ring of circumference 
$L$ with Hamiltonian \cite{Sut71}
$-\sum_{j=1}^N\frac{\partial^2}{\partial x_j^2}
+g\sum_{1\le j<k\le N}d^{-2}(x_{jk})$ in units with $\hbar=2m=1$,
where $d(x_{jk})=(L/\pi)\sin(\pi|x_j-x_k|/L)$ 
is the chordal distance between $x_j$ and $x_k$. It may be realizable via the 
strong dipolar interactions in $^{52}$Cr \cite{Yu06,Gri05}.
The same Hamiltonian can be applied to wave functions with anyonic
symmetry, and the anyonic ground state obtained by mapping from the
fermionic one \cite{Sut71}: 
$\Psi_{A0}(x_{1},\cdots,x_{N})=
\prod_{1\le j<k\le N}\epsilon(x_{jk})e^{-\frac{1}{2}i\theta\epsilon(x_{jk})}
|\sin(\pi x_{jk}/L)|^\lambda$ with
$\lambda=\frac{1}{2}(1+\sqrt{1+2g})$ and energy
$E_0=\frac{1}{3}\pi^2\lambda^2 N(N^2-1)/L^2$, reducing for $g=0$ to the
corresponding results for the TG gas \cite{Gir60}. 

{\it Anyonic TG gas:} The TG gas is a 1D gas of impenetrable point bosons
($a\to 0+$ limit of hard cores of diameter $a$) with no particle-particle
interactions except for the zero-diameter hard cores, which are equivalent
to a constraint that all wave functions vanish at particle collision points
$x_j=x_k$. It has been solved exactly by FB mapping to the
ideal Fermi gas for the cases of periodic boundary conditions \cite{Gir60},
harmonic trapping \cite{GirWriTri01}, and box enclosure \cite{dCamMug05}.
Consider now an anyon gas. In the absence of interparticle
interactions such a system is sometimes called an ideal anyon gas, but
since (\ref{symmetry}) implies that its wave functions $\Psi_A$ automatically
satisfy the impenetrable point constraint of vanishing at contacts
$x_j=x_k$, it is more properly viewed as an anyonic generalization of the
TG gas, hence the name ``anyonic TG gas''. It maps to
the 1D ideal Fermi gas via 
$\Psi_A(x_1,\cdots,x_N)=A_{-\theta}\Psi_F(x_1,\cdots,x_N)$. The
Schr\"{o}dinger equation is to be applied only when all interparticle
separations $|x_j-x_k|$ are nonzero, being replaced at particle contact
$x_j=x_k$ by the condition of vanishing wave function. The Hamiltonian
consists only of the kinetic energy operator, and it follows just as for the
TG gas \cite{Gir60,Gir65} that its complete energy spectrum 
is identical with that of the corresponding ideal Fermi gas, as are all
properties (both time-independent and time-dependent \cite{YukGir05})
depending only on absolute squares $|\Psi_A|^2$ of its wave functions.
For twisted boundary conditions its ground state wave function is
$\Psi_{A0}=\prod_{1\le j<k\le N}\sin(\pi x_{jk}/L)
e^{-\frac{1}{2}i\theta\epsilon(x_{jk})}$
where $x_{jk}=x_j-x_k$, its energy in the thermodynamic 
limit is the same as that of the TG gas, $E_0/N=(\pi\hbar n)^2/6m$, and
its low excitation spectrum is of phonon form with sound speed 
$c=\pi\hbar n/m$ \cite{Gir60}; here $n=N/L$, the particle number density.
These thermodynamic limit results also apply to the anyonic TG gas on a
ring of circumference $L$, since, as previously pointed out, the values
of $\theta$ allowed by the requirement of single-valued wave functions
become dense in the thermodynamic limit. 

{\it Anyonic FTG gas:} The FTG gas is a spin-aligned 1D Fermi gas with
infinitely strongly attractive zero-range odd-wave interaction induced by
a p-wave Feshbach resonance. It is the infinite 1D scattering
length limit $a_{1D}\to -\infty$ of a 1D Fermi gas with zero-range
attractive interactions leading to a 1D scattering length defined 
in terms of the ratio of the derivative $\Psi_{F}^{'}$ of the wave function
to its value at contact: 
$\Psi_{F}(x_{jk}=0+)=-\Psi_{F}(x_{jk}=0-)
= -a_{1D}\Psi_{F}^{'}(x_{jk}=0\pm)$ \cite{GraBlu04,GirOls03,GirNguOls04}.
The limit $a_{1D}\to -\infty$ corresponds to a zero-energy scattering
resonance reachable by Feshbach resonance tuning to a 1D 
confinement-induced resonance \cite{Rob01,Ols98,BerMooOls03,GraBlu04}, 
where the exterior 1D scattering wave function is constant. 
The contact discontinuities of $\Psi_F$ \cite{CheShi98} can be understood as a
zero-range limit $x_0\to 0+$ and $V_0\to\infty$ of the two-body solution for 
a square well of width $2x_0$ and depth $V_0$, where the limit is 
carried out such that $V_0 x_0^2$ approaches a finite, nonzero limit 
\cite{GirOls03,GirNguOls04}. For $a_{1D}\to -\infty$ the exterior solution
is constant (+1 for $x_{12}>0$ and -1 for $x_{12}<0$) and the interior
solution is $\sin(\kappa x_{12})$ with $\kappa=\sqrt{mV_{0}/\hbar^2}=\pi/2x_0$.
In the zero-range limit the interior kinetic energy $\to +\infty$ and
potential energy $\to -\infty$, but their sum remains zero, the ground state
energy. The $N$-body problem is solved by mapping to the ideal Bose gas
\cite{GirOls03,GirNguOls04}. For periodic boundary conditions this Bose
ground state is a trivial constant. However, by looking at the 
square well solution one sees that there is a nontrivial interior Bose
wave function $\sin(\kappa |x_{12}|)$ vanishing with a cusp at $x_{12}=0$.
Therefore, physical consistency requires the presence of a zero-diameter
hard core interaction added to the square well. The mapped Bose
gas is then not truly ideal, but rather a \emph{TG gas with 
superimposed attractive well}, whose nontrivial interior wave function becomes 
invisible in the zero-range limit, simulating an ideal Bose gas 
insofar as the energy and exterior wave function are concerned. The required
impenetrable core is physically quite reasonable, since the atoms have a strong
short-range Pauli exclusion repulsion of their inner shells, whose diameter is 
effectively zero and strength infinite on length and energy scales appropriate 
to ultracold gas experiments.  

This approach is easily generalized to the anyonic case. In the resonant case
$a_{1D}\to -\infty$ the exterior two-body wave function is 
$\epsilon(x_{12})e^{-\frac{1}{2}i\theta\epsilon(x_{12})}$ and the interior
wave function is $e^{-\frac{1}{2}i\theta\epsilon(x_{12})}\sin(\kappa x_{12})$
with the same value of $\kappa$ as for the FTG gas. This generalizes to 
$N>2$ giving an almost trivial ground state: 
$\Psi_{A0}(x_{1},\cdots,x_{N})=
\prod_{1\le j<k\le N}\epsilon(x_{jk})e^{-\frac{1}{2}i\theta\epsilon(x_{jk})}
\prod_{j=1}^{N}\phi_{0}(x_{j})$ where $\phi_0$ is the lowest ideal Bose
gas orbital for given boundary conditions. The ground state
energy and all properties depending only on $|\Psi_{A0}|^2$ are the same as 
those of the ideal Bose gas. 

A very interesting odd/even $N$ effect occurs if the system is contained on a 
ring of circumference $L$. For the FTG gas we found \cite{GirMin06} that for 
even $N$ the mapped bosonic ground state $\Psi_{B0}$ 
is an antiperiodic BEC fragmented between the $k$-space sites
$k=\pm\pi/L$, thus forcing periodicity of $\Psi_{F0}$ in
view of the antiperiodicity of the mapping for even $N$. However,
for anyonic symmetry one can instead choose the lower-energy 
Bose ground state with all particles condensed at $k=0$, and 
force periodicity by requiring that $\theta$ be an odd integral 
multiple of $\pi/(N-1)$. For $N=2$ and $\theta=\pi$, $\Psi_{A0}$ reduces
to a ``bosonic FTG gas'', but for $N\ge 4$ there are true zero-energy anyonic 
ground states of the form
$\Psi_{A0}
=\prod_{1\le j<k\le N}\epsilon(x_{jk})e^{-\frac{1}{2}i\theta\epsilon(x_{jk})}$.
For even $N$ choosing $\theta$ to be an odd integral multiple of $=\pi/(N-1)$ 
avoids discontinuities at boundaries between adjacent periodicity cells 
\cite{Note1}. From this one proves \emph{Theorem 1: For even $N\ge 4$ there is 
a single-valued, continuous, and periodic anyonic ground state of this system 
on a ring, with anyonic phase slips which
are odd integral multiples of $\pi/(N-1)$, with energy lower by an amount 
$Nh^2/8mL^2$ than that of the fermionic ground state with the same 
interaction.} 

{\it Spinor fermions with anyonic spatial symmetry:} Transitions from 
spin-aligned fermion states to anyonic states cannot be generated by
realistic interactions, but by generalizing to a spinor (spin-free) Fermi gas
one can obtain a gas of spin-$\frac{1}{2}$ fermions with anyonic 
\emph{spatial} symmetry by generalizing the previous state $\Psi_{A0}$ to
$\Phi_{F0}=\prod_{1\le j<k\le N}\epsilon(x_{jk})
e^{-\frac{1}{2}i\theta\epsilon(x_{jk})(\delta_{\sigma_j \uparrow}
\delta_{\sigma_k \downarrow}
-\delta_{\sigma_j \downarrow}\delta_{\sigma_k \uparrow})}$.
$\Phi_{F0}$ is fermionic under combined space-spin exchange,
but undergoes a fermionic sign change plus a phase slip $\pm \theta$ 
under exchange of only space coordinates 
($x_j,x_k$) if $\sigma_j\ne\sigma_k$, but only the fermionic sign change
if $\sigma_j=\sigma_k$. In general there are both even and odd-wave 
interactions, in which case the exact ground state will not have 
this simple form \cite{GirOls04-1,GirNguOls04}, but if the even-wave
repulsion is weak (dimensionless even-wave coupling constant
$\gamma_e\ll 1$) and odd-wave attraction infinite as in the FTG gas
($\gamma_o\to\infty$), 
then the ground state will be approximately of this form. 
$\Phi_{F0}$ is an \emph{exact energy eigenstate of energy zero} if 
$\gamma_e=0$ and under the FB mapping it maps to the ideal Bose gas ground 
state which is totally condensed at $k=0$, whereas the degenerate spin-aligned 
FTG ground states on a ring map to an antiperiodic Bose condensate fragmented 
between $k=\pm\pi/L$ \cite{GirMin06}. $\Phi_{F0}$ is connected to these
spin-aligned states by dipolar interactions, enabling energy lowering by
spin flips and anyonic phase slips.  
$\Phi_{F0}$ is not an exact eigenstate of $\hat{S}_z$, and in fact the 
ground state is degenerate with respect to both $S_z$ and total spin $S$
along the hyperbola $\gamma_e\gamma_0=4$ in the ($\gamma_e,\gamma_o$) 
plane; see pp. 19,20 of 
\cite{GirNguOls04}. This is consistent with our assumptions  
$\gamma_e\to 0$ and $\gamma_o\to\infty$ if the limits are taken along this 
line. $\Phi_{F0}$ has $<\hat{S}_z>=0$, and furthermore, the 
projected state $\Psi_{F0}=\hat{P}_0\Phi_{F0}$ is
an eigenstate of $\hat{S}_z$ with eigenvalue zero and the same (zero) energy,
assuming no spin-dependent interactions; here 
$\hat{P}_0=(2\pi)^{-1}\int_{0}^{2\pi}d\phi\ e^{i\phi\hat{S}_z}$ is the
$S_z=0$ projector. Then $\Psi_{F0}$ satisfies periodicity exactly if 
$\theta$ is chosen to be an odd multiple of $2\pi/N$ \cite{Note2}, implying
\emph{Theorem 2: $\Psi_{F0}$ is an exact ground state of the spinor FTG gas 
with even $N$ on a ring if $\theta$ is an odd multiple of $2\pi/N$. It lies 
lower than the lowest spin-aligned state by an amount $Nh^2/8mL^2$}.

I close by pointing out that $\Phi_{F0}$ has superconductive ODLRO as in 
\cite{GirMin06}, but now for $(x_1,\uparrow;x_2,\downarrow)$ pairs.
\begin{acknowledgments}
I thank Ewan Wright, Maxim Olshanii, Anna Minguzzi, and Brian Granger for 
helpful suggestions. This research was supported by 
U.S. Office of Naval Research grant N00014-03-1-0427 through a 
subcontract from the University of Southern California.
\end{acknowledgments}
\end{document}